# Reliable lift-off patterning of graphene dispersions for humidity sensors

*Jorge Eduardo Adatti Estévez, Fabian Hecht, Sebastian Wittmann, Simon Sawallich, Annika Weber, Caterina Travan, Franz Hopperdietzl, Ulrich Krumbein, Max Christian Lemme\**

*J. E. Adatti Estévez, F. Hecht, A. Weber, C. Travan, U. Krumbein*
*Infineon Technologies AG*
*Am Campeon 6, 85579 Neubiberg, Germany*

*J. E. Adatti Estévez, S. Wittmann, S. Sawallich, F. Hopperdietzl, M. C. Lemme*
*Chair of Electronic Devices, RWTH Aachen University*
*Otto-Blumenthal-Straße 2, 52074 Aachen, Germany*
*Email: max.lemme@eld.rwth-aachen.de*

*S. Wittmann, F. Hopperdietzl*
*Infineon Technologies AG*
*Wernerwerkstraße 2, 93049 Regensburg, Germany*

*S. Sawallich*
*Protemics GmbH*
*Otto-Blumenthal-Straße 25, 52074 Aachen, Germany*

*M. C. Lemme*
*AMO GmbH*
*Otto-Blumenthal-Straße 25, 52074 Aachen, Germany*

Dispersion-based graphene materials are promising candidates for various sensing applications. They offer the advantage of relatively simple and fast deposition via spin-coating, Langmuir-Blodgett deposition, or inkjet printing. Film uniformity and reproducibility remain challenging in all of these deposition methods. Here, we demonstrate, characterize, and successfully apply a scalable structuring method for graphene dispersions. The method is based on a standard lift-off process, is simple to implement, and increases the film uniformity of graphene devices. It is also compatible with standard semiconductor manufacturing methods. We investigate two different graphene dispersions via Raman spectroscopy and Atomic Force Microscopy and observe no degradation of the material properties by the structuring process. Furthermore, we achieve high uniformity of the structured patterns and homogeneous graphene flake distribution. Electrical characterizations show reproducible sheet resistance values correlating with material quantity and uniformity. Finally, repeatable humidity sensing is demonstrated with van der Pauw devices, with sensing limits of less than 1% relative humidity.

## 1. Introduction

Environmental sensing is a field that receives increased attention, which drives a demand for real-time detection and monitoring of different ambient parameters. Solid-state gas sensors are suitable candidates to fulfill these requirements due to their small sizes, high sensitivity toward humidity and a wide range of gases, and low cost.[1,2] In this regard, chemiresistive graphene sensors for measuring humidity and hazardous gases offer noteworthy advantages due to graphene's two-dimensional nature.[3–5] Graphene provides large surface-to-volume ratios, high electrical conductivity, and low noise levels, which altogether constitute beneficial conditions for gas sensing.[6–9] Dispersion-based graphene,[10–12] in particular, is interesting due to low-cost and scalable device fabrication via different deposition methods, such as inkjet printing[13–15], spin-coating[7,8,16,17] or Langmuir-Blodgett[18–20] deposition. Nevertheless, different deposition methods are accompanied by film structure and homogeneity challenges, which translate into device performance and reproducibility issues.[13,17,21]

Inkjet printing offers a fast, easy, patterning-free, and up-scalable deposition technique.[22–24] Nonetheless, it usually requires a significant effort towards optimizing the dispersions' rheological properties[25–27], thus limiting its application. Even if the dispersion printability is optimized, the so-called coffee ring effect is detrimental to sensor repeatability and functionality.[13] Therefore, spin-coating[28] (SC) and Langmuir-Blodgett deposition[29] constitute attractive alternatives to achieve highly uniform graphene thin films. SC, in particular, is an easily applicable and well-established method in the semiconductor industry and can offer a flexible and fast platform for the deposition of different materials, such as polymers, photoresists, or colloidal fluids. Yet, its large-area and non-selective deposition makes a subsequent structuring of the graphene film unavoidable.

In this work, we present a scalable and automatable structuring process for producing uniform and thin dispersion-based graphene films, which we prove to be applicable to different graphene materials and solvents. It relies on two extensively employed methods in the semiconductor industry: photolithography for defining device layouts and, upon graphene deposition, the standard lift off[30,31] technique for pattern definition. The present method differentiates itself from similar works dealing with lift-off patterning of dispersion-based graphene[32,33] in the reduced number of process steps, the resulting simplicity of implementation, and, in particular, compatibility with standard semiconductor technology. We show that our patterning method does not deteriorate the material properties of the graphene and validate the achieved film uniformity and sheet resistance reproducibility by electrical measurements. Finally, we demonstrate the application of the developed method by fabricating humidity sensors with high signal-to-noise ratios.

## 2. Results and discussion
### 2.1. Sample and device fabrication

**Figure 1** illustrates the lift-off patterning process presented in this work. It is briefly described here, while process details can be found in the methods section. The first step is to prepare the substrates with the lift-off pattern. For this, the positive photoresist is deposited by SC (**Figure 1a**), and the desired resist structures are defined via optical lithography on thermally grown $SiO_2$-on-Si substrates (**Figure 1b**). In the next step, the samples are exposed to oxygen ($O_2$) plasma (**Figure 1c**) to precondition the surface for the subsequent deposition of graphene dispersion. Next, the graphene dispersion is deposited via SC onto the surface (**Figure 1d**). SC parameters are chosen to enable graphene dispersions to form a conducting film. Finally, the photoresist layer is removed through a cleaning step of acetone, assisted by either ultra-sonication (US) or manual stirring, followed by an isopropanol bath (**Figure 1e**). This fabrication process produces patterned conductive graphene flake films (**Figure 1e**). We mainly employed a water-based graphene dispersion (G-$H_2$O-Disp) and a terpineol-based graphene dispersion (G-Org-Disp). However, several experiments were conducted with a different version of G-Org-Disp based on the low boiling point solvent isopropanol (G-Org-Disp-lb).

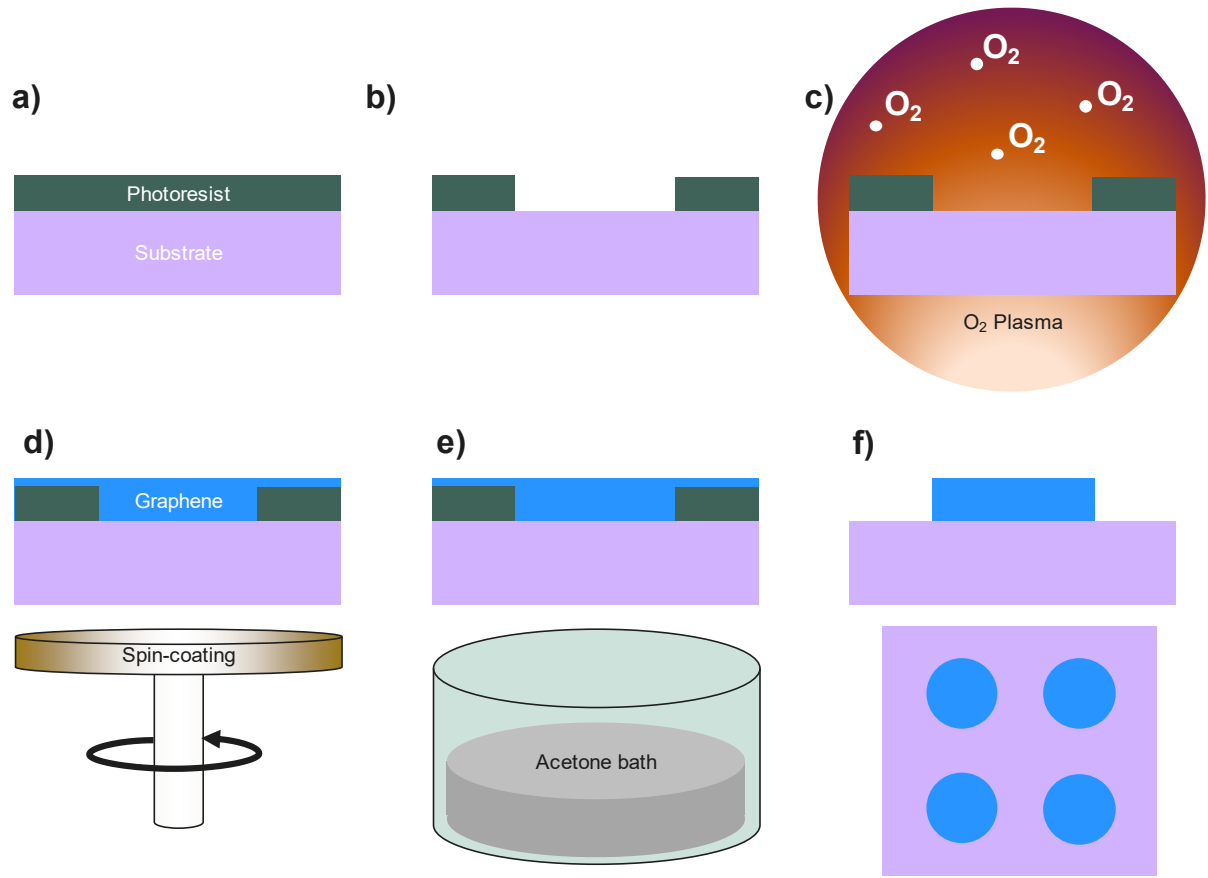

**Figure 1.** Schematic illustration of lift-off patterning process. a) Photoresist is deposited on substrate via spin-coating. b) Pattern on photoresist is defined using optical lithography. c) $O_2$-plasma exposure preconditions surface for the subsequent deposition of graphene dispersion. d) Graphene dispersion is deposited on surface via spin-coating. e) Acetone bath, assisted either by ultra-sonication or manual stirring, removes photoresist and graphene on top of it. f) Cross section (top) and top view (bottom) illustrate resulting graphene sample after completion of lift-off patterning.

## 2.2. Sample definition and summary of analytical methods

The experiments were conducted on different types of samples. **Figure 2a** shows unstructured spin-coated samples (Type-A). They were used for sheet resistance ($R_{Sheet}$) measurements by the four-point probe (4PP) method and Terahertz Time Domain spectroscopy (TDS-THz). With 4PP measurements, we evaluated the effect of the acetone cleaning step on the electrical resistance of spin-coated graphene samples. The TDS-THz measurements aimed to characterize THz-TDS sheet resistance ($R_{Sheet\_THz}$) on both as-coated graphene dispersions (G-H$_2$O-Disp, G-Org-Disp). Type-B samples (**Figure 2b**) consist of lift-off patterned round dots of graphene dispersion. We used them to evaluate the two lift-off process parameters, $O_2$ plasma and US, by optical microscopy and to conduct Raman measurements. On Van der Pauw (vdP) structures (Type-C, **Figure 2c**), we measured the uniformity of the patterned structure by Atomic Force

Microscopy (AFM), and evaluated the variability of vdP sheet resistances ($R_{Sheet\_vdP}$). Furthermore, Type-C structures were selected as humidity-sensing devices. Type-D and Type-E structures (**Figure 2d** and **Figure 2e**) correspond to long-range and short-range transmission line model (TLM) structures, with channel lengths/distances between contacts of $L_{ch}$ = 110 μm to 710 μm and $L_{ch}$ 7 μm to 200 μm, respectively. They were used to assess the variability of TLM sheet resistance ($R_{Sheet\_TLM}$) by DC resistance measurements. On Type-C, Type-D, and Type-E samples, the lift-off patterning described in **Figure 1** was performed on samples that had been previously metalized. Details of the metal contacts are provided in the Methods section. The flowchart in **Figure 2f** provides an overview of the aspects of lift-off patterning analyzed, including the analytical methods applied and sample type investigated.

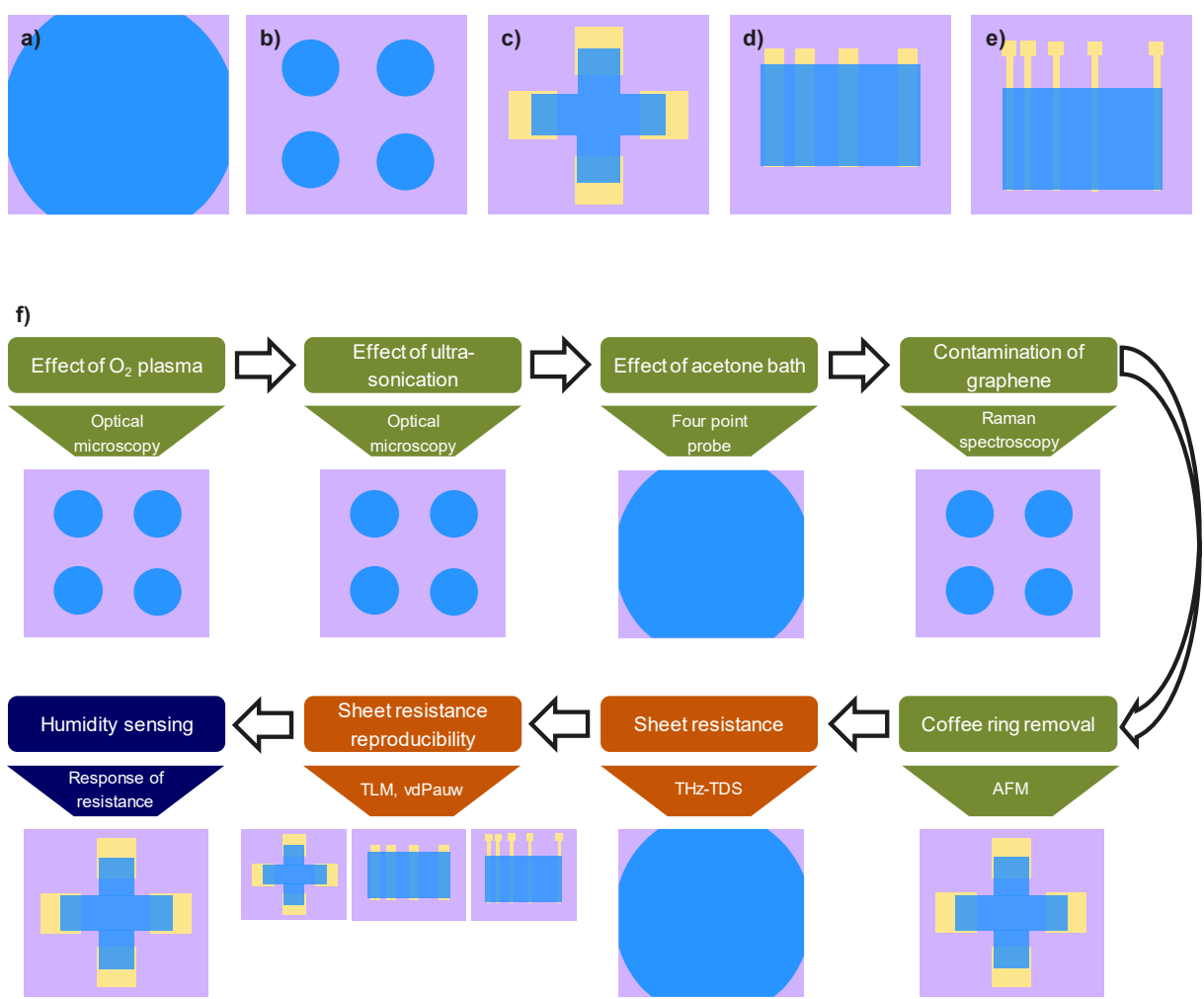

**Figure 2.** Schematic overview of the sample types investigated in this work (purple: substrate, blue: graphene, yellow: metal contacts), and flowchart summarizing the analyses performed in this work. a) Type-A samples consisted of graphene dispersion spin-coated onto a wafer. No patterning was performed. b) Type-B samples contained lift-off patterned round dots of graphene. c) Type-C samples were lift-off patterned Van der Pauw samples. d) Lift-off

patterned long-range TLM samples (Type-D) enabled the measurement of electrical resistances at contact distances between 110 µm and 710 µm. e) On lift-off patterned short-range TLM samples (Type-E), resistances at contact distances between 6 µm and 200 µm were measured. f) The flowchart summarizes the sequence of process, material, and device aspects analyzed in this work. Additionally, it provides information on the analysis method and the sample type used. The analyses performed cover three thematic categories: Process evaluation (green), Evaluation of sheet resistance and sheet resistance reproducibility (orange), and Device application (blue).

## 2.3. Process evaluation

*2.3.1. Effects of $O_2$ plasma*

As a first experiment, we explored the influence of $O_2$ plasma exposure on the patterning process with G-H$_2$O-Disp and G-Org-Disp. Type-B (**Figure 2b**) samples were used for this analysis. In the case of G-H$_2$O-Disp, the lack of a plasma pre-treatment prevents the dispersion from spreading over the surface (**Figure 3a**) and leads to a strong material agglomeration upon applying the lift-off process (**Figure 3b**). The plasma treatment causes surface hydrophilization[34–36], and leads to a homogeneous distribution of the water-based dispersion (**Figure 3c**), which enables the patterning of uniform structures (**Figure 3d**). **Figure 3e** shows the swelling of the photoresist due to terpineol. **Figure 3f** confirms that lift-off without plasma results in an uncontrollable process, evidenced by the graphene dispersion outside of the desired patterns. Therefore, we rely on plasma (**Figure 3g**) to accomplish photoresist hardening[37], which also prevents the organic solvents of G-Org-Disp and G-Org-Disp-lb from dissolving the sacrificial photoresist. This results in a uniform structure after the lift-off (**Figure 3h**).

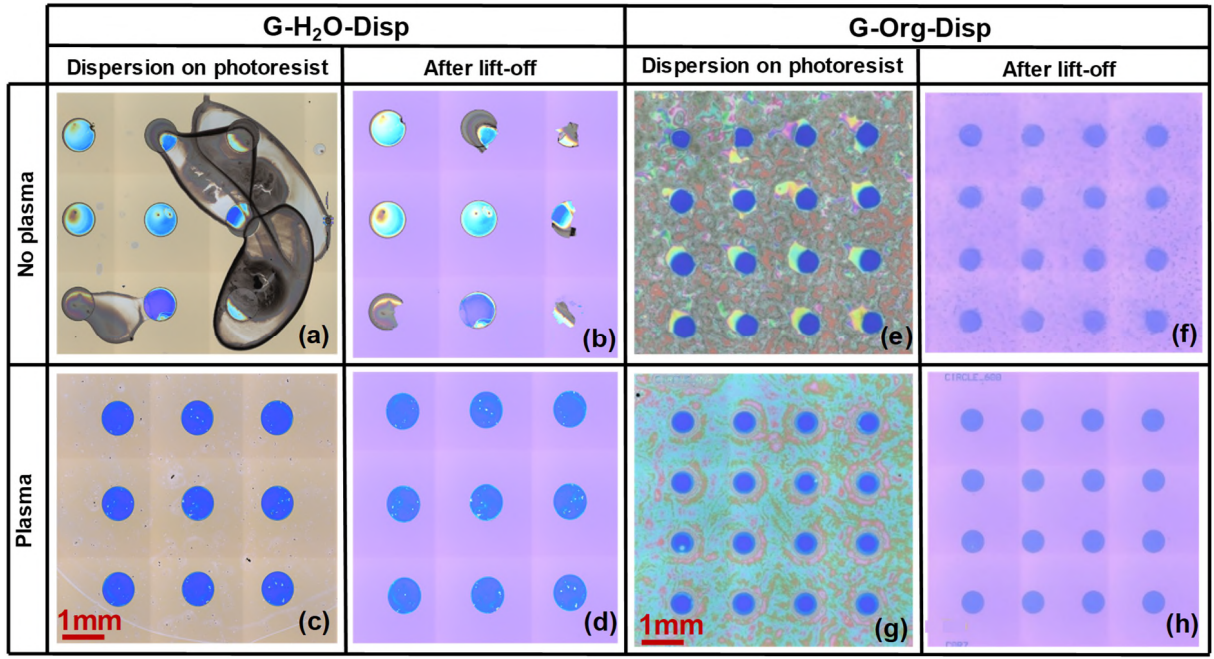

**Figure 3.** Optical investigation of the effect of $O_2$ plasma on the patterning process with G-$H_2O$-Disp and G-Org-Disp. The *"Dispersion on photoresist"* columns show Type-B samples as spin-coated on top of the patterned photoresist. The *"After lift-off"* columns correspond to the same samples after photoresist removal. This comparison is provided for the case with and without plasma to highlight the effect of $O_2$ plasma exposure on the photoresist and the resulting graphene patterning. a) Poor spreading of G-$H_2O$-Disp on an untreated sample. The dispersion tends to agglomerate on photoresist-free regions (circles) and is not evenly distributed on the photoresist surface. b) Structured sample after lift-off with a lack of uniformity and partial removal of strongly agglomerated graphene. c) Enhanced spreading of G-$H_2O$-Disp on both surfaces ($Si_2O$ and photoresist) after $O_2$-plasma treatment. d) Patterned sample after lift-off with a uniform distribution of graphene flakes and a clear definition of the desired circular graphene patterns. e) Photoresist swelling after coating with G-Org-Disp due to the effect of the organic solvent terpineol on the photoresist. f) Structured sample after lift-off, flakes lying outside desired patterns as a result of an incontrollable process due to photoresist swelling. g) Enhanced spreading of G-Org-Disp on hardened photoresist after plasma treatment. h) Patterned sample after lift-off with a clear definition and uniform distribution of graphene flakes within the desired patterns.

The plasma exposure time after the lithographic step directly affects the result of our process and was therefore optimized. We prepared seven samples with G-$H_2O$-Disp and G-Org-Disp-lb, exposed them to plasma for 0s to 30s (in 5s steps), and tracked the result of the lift-off patterning using optical micrographs. Furthermore, the software *Image J*[38] was used to measure the areal deviation between lithographically defined patterns and the resulting graphene structures ($\Delta_{area}$) to quantify the patterning effectiveness. The corresponding percentage values are presented in the bottom right corner of each subfigure in **Figure 4**. **Figure 4a** to **Figure 4g** correspond to G-Org-Disp-lb, while **Figure 4h** to **Figure 4n** show the G-H$_2$O-Disp samples. In **Figure 4a**, isopropanol has penetrated the photoresist, rendering it impossible to remove, similar to the behavior observed with terpineol (**Figure 3e**). According to **Figure 4b**, 5 s are already sufficient to enable the removal of the photoresist. Nevertheless, the image is characterized by the presence of graphene material in the immediate surroundings of the circular structures, suggesting a lower degree of photoresist hardening at the edges. This effect is reduced by increasing the plasma duration to 10 s and 15 s (**Figure 4c** and **Figure 4d**), which reduces the areal deviation to $\Delta_{area}$ = 26.9 % and $\Delta_{area}$ = 33 %, respectively. However, a fully hardened photoresist with resulting low areal deviations $\Delta_{area}$ = 5.3 % and $\Delta_{area}$ = 3.2 % is achieved only after 20 s and 25 s (**Figure 4e** and **Figure 4f**), respectively. **Figure 4g** shows the effect of 30 s plasma and the resulting excessive photoresist hardening, rendering it more challenging to remove and thus causing a less effective patterning. In summary, this experiment shows that 25 s is the optimum duration for G-Org-Disp.

In the case of G-H$_2$O-Disp, decisively different behavior is expected based on the results of **Figure 3b**, where the dispersion tends to agglomerate without plasma pre-treatment, especially at the regions free of photoresist (see also **Figure 4h**). The images suggest that surface hydrophilization could be a potential remedy. To this end, even a short plasma exposure of 5s proved sufficient and enabled an efficient patterning with $\Delta_{area}$ = 3.2 % (**Figure 4i**). Similar low deviations between $\Delta_{area}$ = 1 % and $\Delta_{area}$ = 3 % were obtained by applying plasma times up to 25 s, as depicted by **Figure 4j** to **Figure 4m**. 30 s of plasma exposure again prevented complete photoresist removal and the structuring of the graphene film (**Figure 4n**).

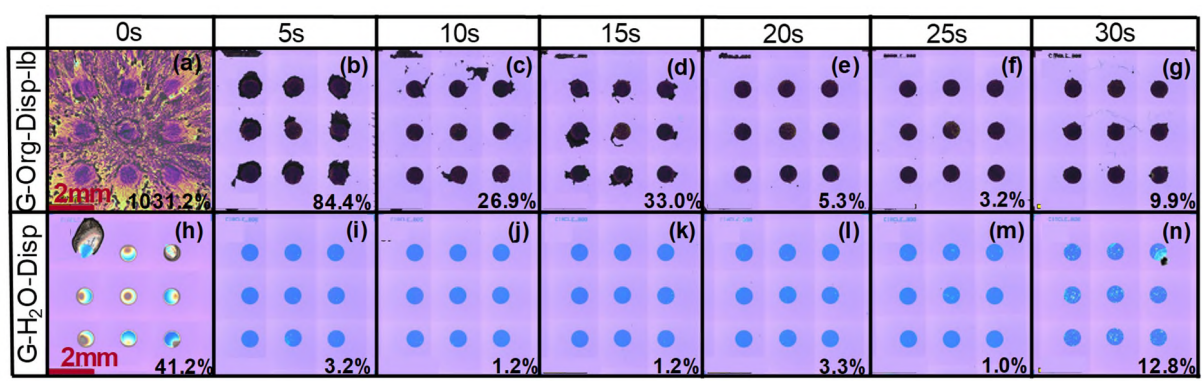

**Figure 4.** Investigation of areal deviation of lift-off structured graphene patterns (numbers in percentage included at the lower right corner of each optical micrograph). a) – g) G-Org-Disp-

lb lift-off patterned after different plasma times (0 s to 30 s in 5 s intervals). Complete photoresist removal is achieved after 20 s and 25 s. Excessive photoresist hardening is observed after 30 s. h) – n) G-H$_2$O-Disp lift-off patterned after different plasma times (0 s to 30 s in 5 s intervals). 5 s are sufficient to ensure full photoresist removal and complete the lift-off. Low areal deviations are achieved for plasma durations between 10 s and 25 s. Excessive photoresist hardening is observed after 30 s.

*2.3.2. Effects of ultra-sonication and acetone bath*

Ultra-sonication is a typical part of lift-off processes to ensure complete removal of the photoresist and increase the repeatability of the patterning result. However, we observed that the mechanical stress applied to the films caused delamination of flakes from the substrate. In the absence of a US step, in contrast, the patterning of G-H$_2$O-Disp remained incomplete, even when we replaced it with a longer acetone bath complemented with manual stirring (**Figure 5a**). We therefore investigated the dispersions' behavior upon US-supported lift-off. We found that a short ultra-sonication (< 5 s) treatment already enabled residual-free structuring of G-H$_2$O-Disp (**Figure 5b**). In the case of G-Org-Disp, the ultra-sonication step led to delamination of the graphene flakes (**Figure 5c** and **Figure 5d**). SEM analysis showed that G-H$_2$O-Disp produces films of flat-lying flakes, while G-Org-Disp contains crumpled flakes with significantly lower flake-to-substrate contact area (**Figure 5e**, **Figure 5g**). We, therefore, mainly attribute the better adhesion of G-H$_2$O-Disp to the flake structure.

We employed the 4PP method to perform a fast and easy assessment of the impact of the acetone bath and the ultra-sonication step on the electrical properties of the graphene films. For this purpose, we used Type-A samples. The acetone stress tests were implemented by exposing the corresponding samples to two runs of a 5-minute acetone bath followed by an isopropanol dip and nitrogen blow drying. On a further G-H$_2$O-Disp sample, we performed a 10s ultra-sonication bath in acetone before the acetone stress test. Ultra-sonication was not applied on G-Org-Disp due to the delamination issue. Subsequently, we tracked the electrical sheet resistance by 4PP ($R_{Sheet\_4PP}$) after different thermal annealing times between 0 and 120 minutes in uncontrolled lab atmosphere. The results of this experiment are presented in **Figure 5f** and **Figure 5h**. The electrical sheet resistance of G-H$_2$O-Disp is neither significantly affected by the acetone bath nor by the US. The resistance values reach their minimum after 10 min anneal at 250°C, regardless of exposure, and remain unchanged with increasing annealing times. This suggests that complete desorption of both acetone and isopropanol from the surface was accomplished. Furthermore, the US did not introduce significant structural defects that degrade the electrical conductivity of the graphene film. In the case of G-Org-Disp, $R_{Sheet\_4PP}$ reaches the same value as a non-exposed sample after 120 minutes of annealing. The thermal annealing efficiently removes acetone, isopropanol, and solvent residuals (**Figure 5f**, **Figure 5h**). We conclude that there is no noteworthy impact of the patterning process on the electrical sheet resistance of either film.

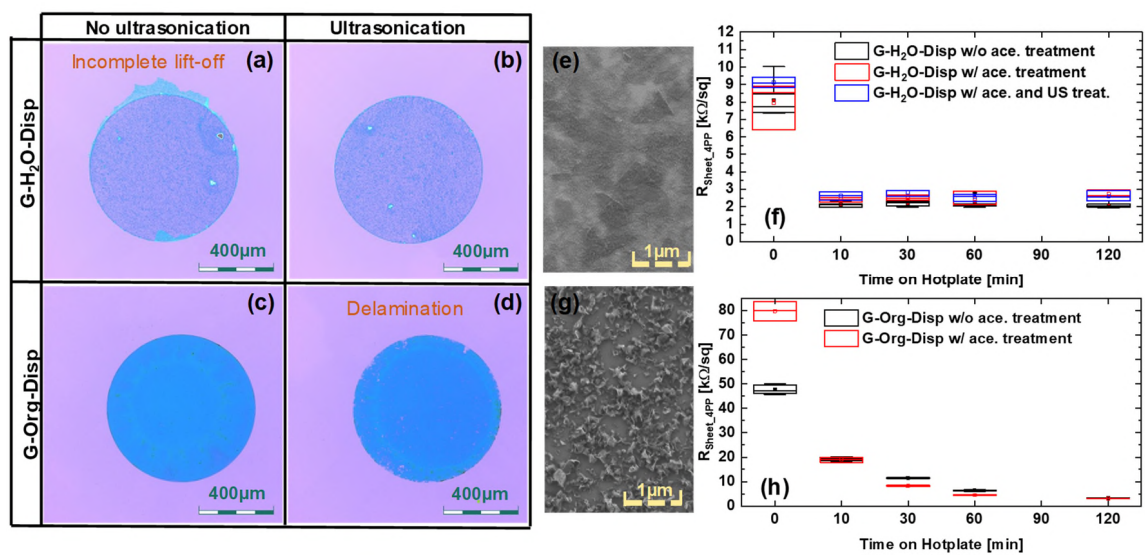

**Figure 5.** Optical analysis of ultra-sonication effect on the structuring process and electrical analysis of the influence of acetone bath and ultra-sonication on the graphene sheet resistance. a) Incomplete structuring of G-H$_2$O-Disp without ultra-sonication. b) Complete structuring of G-H$_2$O-Disp with ultra-sonication. c) Complete structuring of G-Org-Disp without ultra-sonication. d) Delamination of G-Org-Disp upon ultra-sonication. e) SEM micrograph of G-H$_2$O-Disp, flat lying flakes. f) Four-point probe sheet resistance measurements of differently treated samples (Type-A) as a function of annealing time. The resistance minimum was achieved after 10 min anneal regardless of the treatment. g) SEM micrograph of G-Org-Disp crumpled flakes. h) Four probe sheet resistance measurements of differently treated samples (Type-A) as a function of annealing time. Ultra-sonication was not considered due to delamination. The resistance minimum was achieved after 120 min of annealing.

*2.3.3. Impact of lift-off patterning on graphene material properties*

Raman spectroscopy was used to evaluate the impact of the lift-off patterning process on both dispersions, as it constitutes an effective and non-destructive characterization tool for the structure and defect nature of carbon-based materials. In general, the relation between the intensities of the D and G modes ($I_D/I_G$) enables the evaluation of graphene defect density.[39–42]. Therefore, Raman maps were performed on Type-B samples to analyze the areal variability of the Raman fingerprint of our material (see details in Methods section).

**Figure 6a** displays selected Raman spectra of both dispersions and confirms the negligible effect of the structuring process on the material properties of both dispersions. The $I_D/I_G$ ratio of approximately 0.8 in combination with the lack of a prominent 2D-peak confirm a higher defect density and more oxygen-containing functional groups for the water-based dispersion with respect to G-Disp-Org[43,44], which exhibits an $I_D/I_G$ ratio of 0.39 as well as a clear presence of a 2D-peak[45]. The Raman maps on **Figure 6b** confirm the spatial uniformity of both dispersions along the analyzed area. In the case of G-H$_2$O-Disp, we obtained $I_D/I_G$ = 0.8 ± 0.11, regardless of the application of lift-off structuring. For G-Org-Disp, a marginal increase of $I_D/I_G$ from 0.38 ± 0.016 to 0.39 ± 0.018 was observed, but this can be attributed to measurement-to-measurement variations. This confirms that lift-off structuring does not affect the material properties of either dispersion.

*2.3.4. Structural evaluation of lift-off patterned graphene*

Atom Force Microscopy (AFM) offers an important platform to evaluate the resulting structures with respect to their thickness, roughness, and potential undesirable agglomeration patterns at the edges. For this, surface profile measurements as well as topography measurements were performed. G-H$_2$O-Disp (**Figure 6c**) produces a sharp rectangular step and shows no sign of agglomeration at the edges. This corroborates the full removal of photoresist (and graphene on top of it) and, consequently, the effectiveness of the lift-off structuring. In contrast, some agglomeration becomes visible at the edge of the G-Org-Disp structure (**Figure 6e**). This is attributed to the lack of the US step, which leaves graphene flakes on the photoresist positive edges. AFM height (h) and roughness ($S_q$) values confirm the structural differences in the SEM pictures in **Figure 5e** and **Figure 5g**. $h$ = 14 nm and $S_q$ = 4 nm were measured on the G-H$_2$O-Disp sample, while G-Org-Disp showed $h$ = 73 nm and $S_q$ = 33 nm (**Figure 6d**, **Figure 6f**). This confirms the presence of thinner and flatter layers for G-H$_2$O-Disp compared to G-Org-Disp.

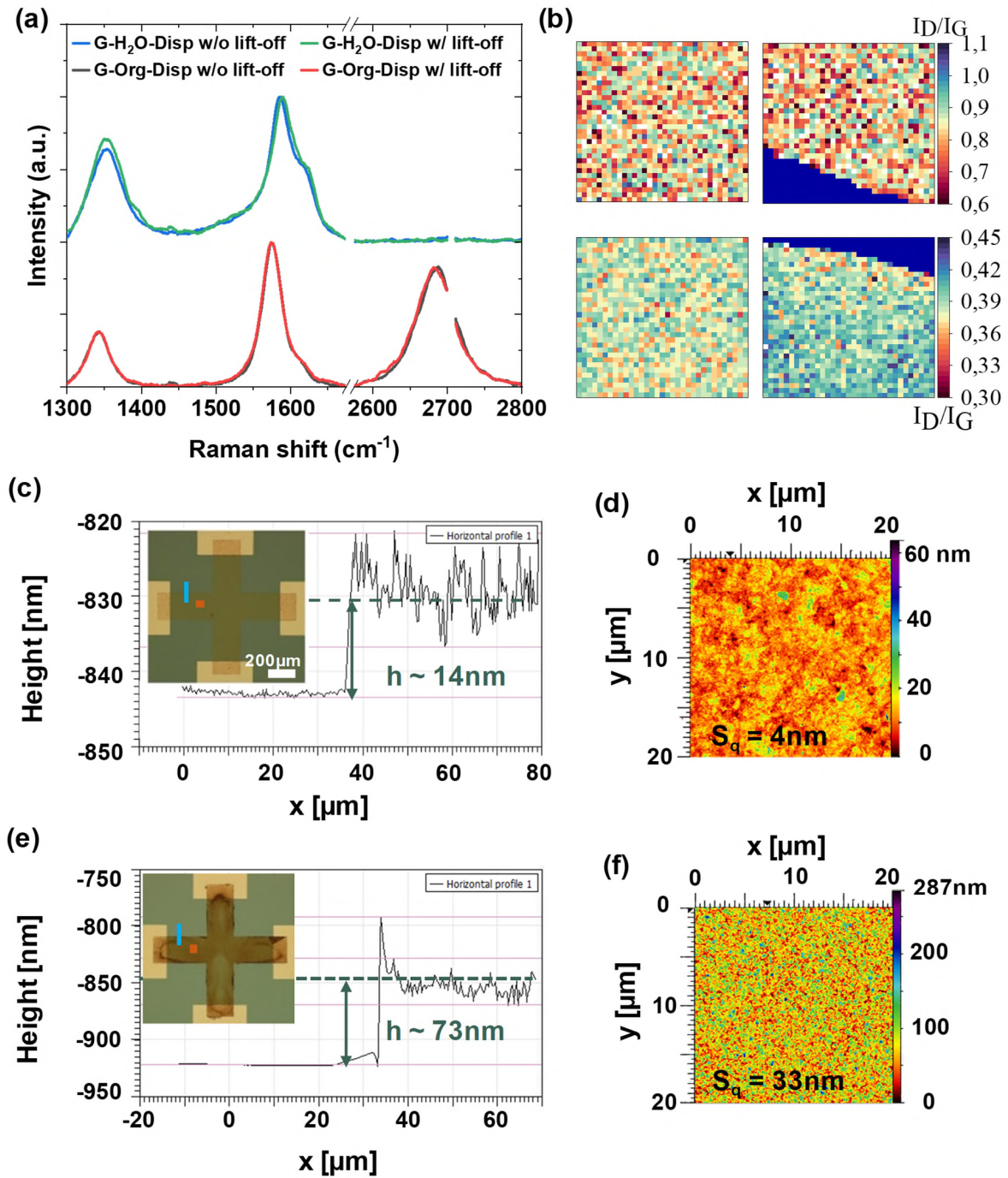

**Figure 6.** Raman and AFM analysis of G-H$_2$O-Disp and G-Org-Disp: a) Single Raman spectra of both dispersions with and without lift-off show no impact of the lift-off process on material properties. Furthermore, they confirm a lower degree of defects of G-Org-Disp compared to G-H$_2$O-Disp. G-Org-Disp exhibits a clear 2D peak and an $I_D/I_G$ ratio of 0.39. G-H$_2$O-Disp does not exhibit a 2D peak and shows an $I_D/I_G$ ratio of 0.8. b) Raman maps of $I_D/I_G$ ratio. Top left: G-H$_2$O-Disp, average $I_D/I_G$ = 0.8; top right: lift-off structured G-H$_2$O-Disp, average $I_D/I_G$ = 0.8; bottom left: G-Org-Disp, average $I_D/I_G$ = 0.38; bottom right: lift-off structured G-Org-Disp, average $I_D/I_G$ = 0.39. c) AFM surface profile of lift-off patterned G-H$_2$O-Disp. Inset: micrograph of Type-C structure including spots for profile (blue) and topography (red)

measurements. No flake agglomeration is visible at the edge of the lift-off patterned structure. The green arrow indicates the extraction of the height value from the profile scan. d) AFM topography scan and roughness measurement of G-H₂O-Disp showing uniform distribution of graphene flakes and the formation of a flat film with $S_q$ = 4 nm. d) AFM surface profile of lift-off patterned G-Org-Disp. Inset: micrograph of Type-C structure including spots for profile (blue) and topography (red) measurements. Flake agglomeration visible at the edge is attributed to incomplete lift-off due to lack of ultra-sonication step. The green arrow indicates the extraction of the height value from the profile scan. f) AFM topography scan and roughness measurement of G-Org-Disp sample showing uniform distribution of graphene flakes and the formation of a film with $S_q$ = 33 nm.

## 2.4. Sheet resistance measurements

### 2.4.1. THz-Time Domain Spectroscopy

THz-TDS has been demonstrated as a versatile tool for semiconductor inspection[46,47] and offers a fast and non-destructive measurement technique to characterize graphene-based materials[48–50]. We applied it to measure the sheet resistance and homogeneity of spin-coated samples of both dispersions. The THz-TDS-scanned samples consisted of Type-A samples with G-H₂O-Disp and G-Org-Disp, spin-coated three times successively. The spatially resolved graphene sheet resistance $R_{Sheet\_THz}$ (x,y) was obtained by mapping the entire samples and by applying the Tinkham formula (**Equation 1**)[51] on the relation between the THz transmission through the substrate and graphene ($T_{SL}$) and the THz transmission through the reference substrate alone ($T_S$),

$$T(\omega) = \frac{T_{SL}(\omega)}{T_S(\omega)} = \frac{1}{1 + \frac{Z_0}{R_{Sheet\_THz}(\omega) \cdot (n_{Si} + 1)}} \quad (1),$$

with the free space impedance $Z_0$ = 377 Ohm and the refractive index of silicon of $n_{si}$ = 3.42.

**Figure 7a** and **Figure 7b** show the $R_{Sheet\_THz}$ maps acquired on G-H₂O-Disp and G-Org-Disp, respectively. These spatially resolved maps allow localizing regions of homogeneous conductivity, i.e. regions most suitable for device fabrication. On G-H₂O-Disp (**Figure 7a**), the different film thicknesses related to the 3 sequentially applied SC layers are inversely correlated with the measured $R_{Sheet\_THz}$. The central sample area contains the targeted three overlapping layers. Here, we measured a considerably lower $R_{Sheet\_THz}$ = 1,6 kOhm/sq compared to $R_{Sheet\_4PP}$ = 9,8 kOhm/sq (section 2.3.2). In contrast, the average values for three SC layers of G-Org-Disp of $R_{Sheet\_THz}$ = 6,83 kOhm/sq and $R_{Sheet\_4PP}$ = 7,5 kOhm/sq lie in a similar range. This difference is explainable by the different carrier movement range and channel lengths achieved with the THz-TDS and 4PP methods[48], as well as the flake structure and size of both dispersions (**Figure 5e**, **Figure 5g**). Firstly, $R_{Sheet\_4PP}$ with a channel length of ~ 200 µm is dominated by interflake resistances and percolative transport[52–54]. Secondly, THz-TDS exhibits short sub-µm interaction lengths, which emphasizes the conductivity of individual flakes. This efficiently probes the low intraflake resistances for G-H$_2$O-Disp, which consists of about 1 µm-sized flat flakes. In the case of G-Org-Disp, characterized by non-planar crumpled flakes with dimensions between 200 and 500 nm, the THz-TDS method results in significantly higher resistances.

*2.4.2. TLM and van der Pauw measurements.*

We measured the DC sheet resistance of lift-off structured samples using transmission line model (TLM) and Van der Pauw (vdP) measurements to assess the uniformity of the patterned structures using Type-D samples (**Figure 7c**). Three Type-D samples were fabricated per dispersion, with 2 to 4 sequential SC runs at 1500 rpm. **Figure 7d** displays a box plot of the obtained sheet resistances of all channel widths $W_{ch}$ from 200 µm to 400 µm in 50 µm steps. The plot shows a decrease in TLM sheet resistance ($R_{Sheet\_TLM}$) proportional to increasing film thicknesses. Thicker films further improve $R_{Sheet\_TLM}$ variability (reduced spread). Both effects can be attributable to enhanced film uniformity[55] when more material is deposited. This leads to a reduction of interflake resistance by shortening the distance between flakes[52] (relevant for crumpled flakes as in G-Org-Disp) and increasing the flake overlap area[56] (relevant for flat-lying, stacking flakes in G-H$_2$O-Disp). The sheet resistances for 3 x 1500 rpm and 4 x 1500 rpm SC are quite uniform and develop linearly up to $L_{ch}$ = 710 µm for both dispersions (see **Figure S1** in Supplementary Information).

From this point on, we focus on G-H$_2$O-Disp due to the dispersion's mechanical robustness to US during lift-off (**Figure 5**) and the higher defect density observed in the Raman measurements (**Figure 6a**), which is expected to be beneficial for humidity sensing[57–59] and confirmed in the *Humidity sensing with both dispersions using TLM devices* section in the Supplementary Information). Further electrical measurements were conducted on Type-C as well as Type-E structures with 2, 3 and 4 SC runs at 1500rpm (see Supplementary Information). **Table 1** provides a summary of all $R_{Sheet}$ measurements performed on both non-patterned (Type-A) and lift-off-patterned (Type-C, Type-D, Type-E) samples. It includes data from the vdP measurements, which are described in detail in the Methods section. The vertical line splits the table in two sections. Each section includes samples fabricated with G-H$_2$O-Disp from different batches (batch 1 and batch 2), which were produced at different lab facilities. We expect this to be the reason for different graphene concentrations leading ultimately to the lower $R_{Sheet\_TLM}$ of batch 1 compared to $R_{Sheet\_TLM}$ of batch 2. The DC resistance measurements by all techniques support the results of the Type-D samples with regard to $R_{Sheet}$ at different thicknesses. Additionally, the excellent comparability between $R_{sheet\_4PP}$ and $R_{Sheet\_TLM}$ (Batch 1) and, especially, between $R_{Sheet\_TLM}$ and $R_{Sheet\_vdP}$ (batch 2) for 3x1500rpm and 4x1500rpm complements the analysis. It demonstrates the reproducibility of the lift-off structured graphene film regardless of DC sheet resistance measurement method and device geometry once a certain thickness (3 x 1500 rpm in our experiments) is reached.

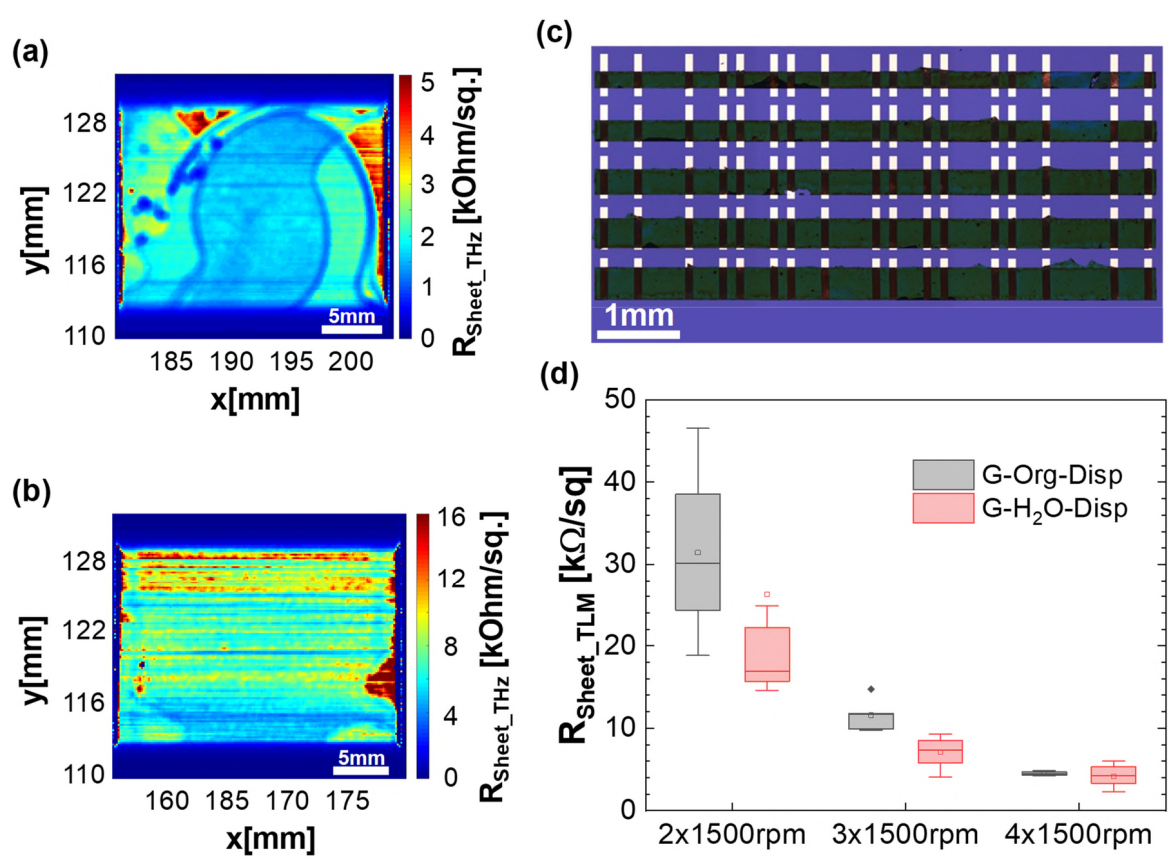

**Figure 7.** Sheet resistance characterization of G-H$_2$O-Disp and G-Org-Disp: a) Sheet resistance measurements on G-H$_2$O-Disp (3 x 1500 rpm) via THz-TDS. Different thicknesses originating from successive spin-coating are visible. The measured sheet resistances correlate inversely with film thicknesses. The central sample area contains three overlapping spin-coating layers. b) Sheet resistance measurements on G-Org-Disp (3 x 1500 rpm) via THz-TDS. The regions at the edges with higher resistance values indicate lower flake coverage. The central sample area corresponds to three overlapping spin-coat layers. c) Micrograph of lift-off-structured TLM-devices for channel lengths from 110 μm to 710 μm in 200 μm steps and channel widths from

200 µm to 400 µm in 50 µm steps (here: G-Org-Disp 4 x 1500 rpm). d) Sheet resistance versus number of spin-coat runs for both dispersions. Increasing the number of spin-coating runs decreased the sheet resistances and improved variability.

**Table 1.** Comparison of sheet resistance measurements on G-H$_2$O-Disp using different methods and dispersion batches. Batch 1 (G-H$_2$O-Disp): 4PP, THz-TDs, and long-range TLM. A good comparability of sheet resistances with 3 x 1500 rpm and 4 x 1500 rpm demonstrates the reproducibility of resistance values. A significant difference between 4PP and THz-TDs (3x1500rpm) measurements is attributed to the measurement techniques: while 4PP captures predominantly interflake resistance, THz-TDS values are dominated by the conductivity of individual flakes. Batch 2 (G-H$_2$O-Disp): short-range TLM and vdP samples. A good comparability of sheet resistance values with 3 x 1500 rpm and 4 x 1500 rpm highlights the homogeneity of lift-off structured graphene films regardless of geometry. Higher deviation between methods at 2 x 1500 rpm evidences a lack of film homogeneity, resulting in lower reproducibility of resistance value.

| Spin-coating runs | $R_{Sheet\_4PP}$ (Type-A) [kOhm/sq] | $R_{Sheet\_THz}$ (Type-A) [kOhm/sq] | $R_{Sheet\_TLM}$ (Type-D) [kOhm/sq] | $R_{Sheet\_TLM}$ (Type-E) [kOhm/sq] | $R_{Sheet\_vdP}$ (Type-C) [kOhm/sq] |
|---|---|---|---|---|---|
| | Batch 1 | | | Batch 2 | |
| 2 x 1500 rpm | 52.6 | | 26.26 | 110.33 | 86.33 |
| 3 x 1500 rpm | 9.84 | 1.6 | 7.08 | 13.98 | 12.98 |
| 4 x 1500 rpm | 5.39 | | 4.17 | 7.12 | 7.19 |

### 2.5. Humidity sensing

Type-C devices were tested at different relative humidity (RH) values between ~25% and 80% at room temperature in a controlled humidity test setup (3 each with 2x1500rpm, 3x1500rpm and 4x1500rpm G-H$_2$O-Disp). **Figure 8a** shows the response of our sensors over time as humidity is changed. We define the response of our sensor as the relative sheet resistance, i.e. $Response = R_{rel} = (R_{Sheet\_vdP} - R_0)/R_0$, with $R_0 = avg(R_{Sheet\_vdP})$ at relative humidity $RH_{min} = 25\%$ and with $R_{Sheet\_vdP}$ as the continuously measured sheet resistance, expressed in percentage for clearer visualization. The exposed devices initially showed an increase of $R_{Sheet\_vdP}$ proportional to relative humidity for all displayed sensors, albeit with some drift. Above 60% relative humidity, the sensors' sheet resistances destabilized further, failing to recover and instead showing higher resistances. Both effects are more substantial with thinner graphene layers. Furthermore, the thicker and more conductive films reduce noise (see details from **Figure 8a** in **Figure S4** of the Supplementary Information for better visualization). We consider two possible explanations for this behavior. Firstly, it has been observed that on multi-layer graphene, the noise arising from the bottom layer represents the dominant noise contribution due to substrate interactions. This noise decreases with an increasing number of layers.[60] A similar behavior has been reported on spin-coated reduced graphene oxide.[7] Secondly, the film homogeneity and conductance are increased by subsequent SC runs by increasing flake overlap, as discussed above. Both aspects are expected to lower the noise levels on percolation films[61], where the interflake resistance plays a significant role.[62] At the same time, thicker graphene films of G-H$_2$O-Disp lead to lower responses of $R_{Sheet\_vdP}$ to RH due to a decreased surface-area-to-volume ratio. We compared the noise density and signal-to-noise ratio (SNR) for 2 x 1500 rpm, 3 x 1500 rpm, and 4 x 1500 rpm, as well as data for 4 x 1500 rpm that included a heating step for humidity desorption (300 s at 110 °C, implemented as explained in the Methods subsection). This step improves the recovery of the $R_{Sheet\_vdP}$ by desorbing H$_2$O[58], and it was carried out for the 4 x 1500 rpm sensors due to their high reproducibility and reduced noise level.

The signal-to-noise ratio is defined as

$$SNR\left(\frac{1}{\%}\right) = \frac{sens}{R_{noise}} \qquad (2)$$

where *sens* is the device sensitivity and $R_{noise}$ is the noise ratio. The sensitivity is defined as the quotient between relative sheet resistance at maximum relative humidity and the difference between maximum and minimum relative humidity at the first humidity pulse (**Figure 8a**), according to **Equation 3**:

$$sens\left(\frac{1}{\%}\right) = \frac{R_{rel\_max} - R_{rel\_min}}{R_{rel\_min} \cdot (RH_{max} - RH_{min})} \qquad (3)$$

$R_{noise}$ is defined as the ratio of the standard deviation of sheet resistance and its averaged absolute value:

$$R_{noise} = \frac{std(R_{Sheet\_vdP})}{avg(R_{Sheet\_vdP})} \qquad (5)$$

We selected for this purpose the portion of the signal corresponding to the 120 s prior to the first humidity pulse, where the measured relative humidity is stable at $RH = 25\ \%$.

The results of this analysis are presented in **Figure 8b**. We measured a decrease of noise amplitude by one order of magnitude with increasing film thicknesses and a somewhat smaller noise spread for the 4 x 1500 rpm devices subjected to the heating step. Furthermore, these latter devices are the only ones that show an SNR ratio clearly above the 3σ level of detection[63] for $\Delta RH = 1\ \%$, indicated by a dashed green line. This corresponds to the limit at which the signal change produced by 1 % change in *RH* equals the threefold of the measured noise at the baseline (*RH* = 25 %). The heating step enhances the SNR far beyond the detection limit (from 3.21 σ to 14.3 σ) for *RH* = 1 %.

**Figure 8c** illustrates the difference between the sensor's response to a humidity exposure directly upon a heating step (left curve) and the sensor response without a heating step (right curve). The exposure of the 4 x 1500 rpm sensor to 110 °C prior to humidity exposure renders it up to 5 times more sensitive compared to the non-heated sample. This relation holds throughout our experiment, in which we measured the sensor response to three successive humidity pulses from 40 % to 80 %, with heating steps in between, demonstrating the repeatability of the response (see Figure S4 in Supplementary Information for further details). This highlights the importance and the impact of refreshing the sensor surface by thermally desorbing water molecules.

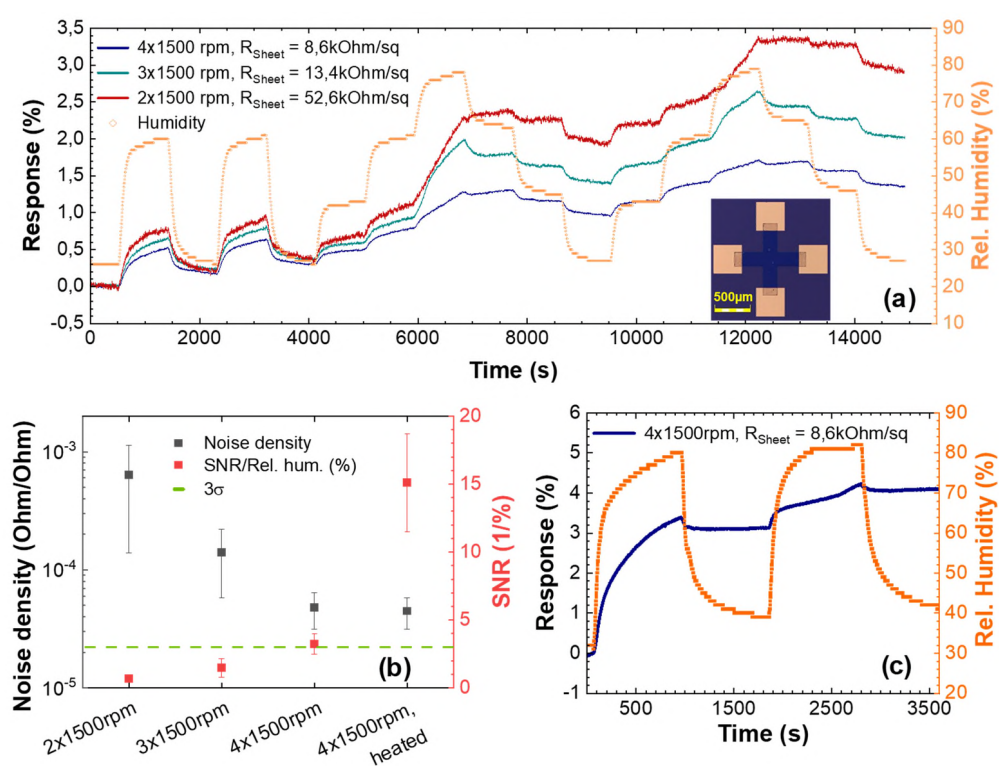

**Figure 8.** Humidity sensor experiments. a) Humidity measurements on Type-C devices of different thicknesses. A clear response to humidity is visible for all devices. The strongest response was achieved for a 2 x 1500 rpm device, however, it also exhibits the most pronounced drift and noise. The lowest drift and noise rate were observed in a 4 x 1500 rpm device. b) Noise density and SNR for different spin-coated layers, including after heating. The noise decreases with increasing film thickness. At 4 x 1500 rpm, the devices achieve a detection limit (3σ) of *RH* = 1 %. This is increased up to five times by implementing a heating step for humidity

desorption. c) The response of a 4 x 1500 rpm device to humidity pulses with and without preceding heating pulses ($R_{Sheet\_vdP\_min}$ at $RH$ = 30 %). The heating pulses render devices up to 5 times more sensitive to humidity due to humidity desorption from the surface.

## 3. Conclusions

We have demonstrated, characterized and successfully applied a scalable, semiconductor technology compatible lift-off fabrication method for humidity sensors based on graphene dispersions. The proposed process yields high film uniformity for different dispersions. Raman and AFM analyses corroborated the negligible impact of the process on the intrinsic material properties as well as the homogeneity of the structures. Moreover, we electrically characterized and compared two different kinds of graphene dispersions by the transmission line model measurement technique as well as by Terahertz Time-Domain spectroscopy. The THz-TDS and TLM data confirm the uniformity of sheet resistance upon spin-coat deposition as well as its correlation with the number of spin-coating cycles for both dispersions. We compared van der Pauw (vdP) and TLM structures with the water-based dispersion G-H$_2$O-Disp and showed reproducible sheet resistances for both device geometries. The water-based dispersion was more robust towards lift-off structuring, as well as more defective, and was shown to be superior for humidity sensing. The thinnest devices with the highest surface-to-volume ratio showed the strongest response to humidity, however, thicker films improved the signal to noise ratio significantly. Finally, we implemented a heating step in the measurement routine to desorb humidity, which enabled detection limits of less than 1% relative humidity and a repeatable response. Our work presents a scalable, semiconductor manufacturing compatible method for graphene device fabrication and its successful application as humidity sensors. Furthermore, it emphasizes the importance of a controlled film formation for chemiresistors to achieve both high electrical resistance reproducibility and low noise amplitudes, resulting in better signal to noise ratios.

## 4. Methods

*Sample fabrication (lift-off patterning process):*

The desired resist structures were defined via optical lithography on thermally grown SiO$_2$ (280 nm thickness) on Si substrates employing a mask aligner MA56 from Karl Süss. For this purpose, we spin-coated a 770 µm thick film of the positive photoresist S1805 from MicroChem onto the substrate, exposed it for 120 s and developed it for 90 s in a metal ion free trimethylammonium-based developer (AZ 726 MIF from MicroChemicals). In a next step, the samples were exposed to O$_2$ plasma using a MyPlas system from Plasma Electronics at 40W and under an O$_2$ flow of 20 sccm. The optimized plasma duration was 25 s.

Subsequently, the graphene dispersion was deposited via SC onto the surface. The selected rotation speed was 1500 rpm. This is the minimum value at which the dispersion is uniformly distributed over the sample and the resulting amount of deposited material enables the formation of a conducting film at the given graphene concentration of 2 g/l. Further SC runs at 1500 rpm were carried out on top of each other to increase the film thickness. Hereupon, the samples were placed onto a hotplate at 110 °C for 60 s to accelerate the drying process of the dispersion.

Finally, the photoresist layer was removed through a cleaning step of acetone for 10 s, assisted by either US (< 5 s) or manual stirring, followed by an isopropanol bath. In order to assure the removing of all solvents as well as residuals from the lift-off process, the samples were placed on a precision hotplate (Harry Gestigkeit GmbH). The optimized parameters were 15 min at 250 °C for G-H$_2$O-Disp and 120 min at 350 °C for G-Org-Disp.

*Sample types:*

All sample types were fabricated on 2.5 cm x 2.5 cm wafer pieces consisting of 280 nm thermally grown SiO$_2$ on Si. The same SC speed of 1500 rpm was used for all samples, and the number of successive SC runs was increased to increase the amount of deposited graphene dispersion.

<u>Type-A</u> samples were spin-coated, non-patterned graphene dispersion samples. The dispersion was always spin-coated after the O$_2$ plasma exposure above. <u>Type-B, Type-C, Type-D and Type-E</u> samples were lift-off patterned according to the above procedure. Furthermore, <u>Type-C, Type-D, and Type-E</u> samples consisted of graphene dispersion patterned on substrates that already had contacts for electrical measurements (see below for more information on metal contacts). <u>Type-B</u> samples consisted of circular dots with graphene dispersion. Each sample contained either 9 (3 x 3) or 16 (4 x 4) dots, depending on their size. <u>Type-C</u> samples consisted of six vdP crosses with a distance of 900 µm between opposing contacts. For their use as humidity sensing devices, they were mounted on a custom printed circuit board (PCB) and wire bonded. <u>Type-D</u> samples were long range TLM bars with variable channel dimensions. The channel length varied from 110 µm to 710 µm in 50 µm increments, and the channel width varied from 200 µm to 400 µm in 50 µm increments. <u>Type-E</u> samples were short range TLM structures with channel lengths ranging from 7 µm to 200 µm and channel widths of 300 µm.

*Metal contacts on Type-C, Type-D and Type-E samples:*

Three types of samples, Type-C, Type-D and Type-E included metal contacts for electrical measurements. In all cases, the metal contacts were evaporated and patterned with standard lift-off prior to the graphene lift-off patterning described above. The metal stack on Type-C and Type-D consisted of 50 nm Au on top of 5 nm Ni. Type-E samples had a Ti/Pt/Au metal stack (bottom to top) with corresponding thicknesses of 50 nm/ 100 nm/ 100 nm.

*Raman Spectroscopy:*

Raman maps were performed over an area of 136 μm x 136 μm with a spot spacing of 4 μm. We used a Horiba Raman system with a laser excitation wavelength of 532 nm and a 100x objective. The data were analyzed using a custom-made software.

*Atom Force Microscopy:*

The AFM system consisted of an NX20 microscope (Park Systems). We performed surface profile measurements with a scan length of 80 μm as well as topography measurements on an area of 20 μm x 20 μm corresponding to 256 x 256 measurement points. Data analysis was performed using the software tool Gwyddion[64].

*Terahertz Time Domain Spectroscopy (THz-TDS):*

We used THz-TDS in transmission mode to measure the sheet resistance and homogeneity of unstructured spin-coated samples with both dispersions. The measurements were taken using a THz near-field imaging setup based on photoconductive micro-probe detectors (TeraCube M2 with TeraSpike TD-800-X-HR-WT, Protemics)[65]. The THz-TDS-scanned samples consisted of two Type-A samples, one with G-$H_2$O-Disp and one with G-Org-Disp. Each of them was spin-coated three times successively.

*Four-point-probe sheet resistance measurement:*

A Jandel RM3-AR four-point probe system allowed the fast and direct measurement of the DC sheet resistance on Type-A samples. A measuring current between 10 μA and 1 mA was set for all measurements.

*DC sheet resistance measurements by TLM and vdP:*

All electrical measurements presented in this work were performed using Keithley 2450 Source Meter units. With exception of the humidity measurements, all electrical measurements were carried out using prober needles. In the case of van der Pauw resistance measurements, $R_{Sheet\_vdP}$ was calculated using the following equation:

$$R_{Sheet\_vdP} = \frac{\pi \cdot V}{\ln(2) \cdot I} \qquad (5)$$

with the current *I*, forced between two opposed contacts, and the voltage *V*, measured between the other pair of opposite contacts. The measurements in section 2.4.2 involved switching along the four different combinations of opposite contacts with a current sweep between $I_{min}$ = -100 µA and $I_{max}$ = 100 µA at each one. The resulting $R_{Sheet\_vdP}$ value corresponds to the average of the four sheet resistances obtained by applying **Equation 5** to each of the IV curves.

The measurements in section 2.5 were conducted without implementing the contact rotation, thus providing a non-interrupted recording of electrical data over time. We selected the samples with the lowest variability between $R_{Sheet}$ at each contact pair combination, and applied **Equation 5** to a single contact combination. The forced constant current was $I$ = 100 µA.

More details on the TLM method and how we applied it in this work can be found in the Supplementary Information.

*Humidity measurements on vdP-devices:*

Type-C structures were employed as humidity sensors. Their electrical sheet resistance ($R_{sheet\_vdP}$) constitutes the sensor measurement quantity and is calculated according to **Equation 5**. Here, we measured $R_{Sheet\_vdP}$ at different humidity levels. The relative humidity was set and controlled by a humidity generator MHG32 from proUmid. We further used an SH25 humidity sensor from Sensirion for reference measurements of the relative humidity inside the custom-designed measurement chamber.

*Heating step on vdP-devices:*

PCB-assembled and wire bonded devices were placed on a Miniware MHP30 mini heating plate to test the effect of a heating step on humidity sensing. The duration of the heating step was 300 s and the set temperature was 110 °C.

**Acknowledgements**

This work has received funding from the European Union's Horizon 2020 research and innovation program under grant agreements 881603 (Graphene Flagship Core3).

# Supplementary Information

# Reliable lift-off patterning of graphene dispersions for humidity sensors

*Jorge Eduardo Adatti Estévez, Fabian Hecht, Sebastian Wittmann, Simon Sawallich, Annika Weber, Caterina Travan, Franz Hopperdietzl, Ulrich Krumbein, Max Christian Lemme\**

*J. E. Adatti Estévez, F. Hecht, A. Weber, C. Travan, U. Krumbein*
*Infineon Technologies AG, Am Campeon 6, 85579 Neubiberg, Germany*

*J. E. Adatti Estévez, S. Wittmann, S. Sawallich, F. Hopperdietzl, M. C. Lemme*
*Chair of Electronic Devices, RWTH Aachen University, Otto-Blumenthal-Straße 25, 52074 Aachen, Germany*

*S. Wittmann, F. Hopperdietzl*
*Infineon Technologies AG, Werner­werkstraße 2, 93049 Regensburg, Germany*

*S. Sawallich*
*Protemics GmbH, Otto-Blumenthal-Straße 25, 52074 Aachen, Germany*

*M. C. Lemme*
*AMO GmbH, Otto-Blumenthal-Straße 25, 52074 Aachen, Germany*
*\*Email: max.lemme@eld.rwth-aachen.de*

**Transmission line measurements (TLM)**

**Long-range TLM** structures consist of lift-off structured dispersion-based graphene films. They exhibit a stripe shape with widths ranging from 200 µm to 400 µm in 50 µm steps. Four different distances between contacts (channel lengths) are measured: 110µm, 310µm, 510µm, 710 µm. The measured resistance $R_{total}$ between each pair of contacts is composed of a channel-length-dependent resistance ($R_{Ch}$) and a contact resistance ($R_C$) according to following equation:

$$R_{total} = R_{Ch} + 2R_C \qquad \text{(eq. S1)}$$

The linear fit of resistance measurements taken at different channel lengths (L) enables the extraction of $R_{sheet}$ (slope multiplied by channel width W) as well as $R_c$ (y-axis intercept divided by two)

$$R_{total} = R_{sheet} \cdot \frac{L}{W} + 2R_C \qquad \text{(eq. S2)}$$

where $R_{Sheet}$ is directly proportional to the distance between contacts. We merged the results from different channel widths by multiplying the total measured resistance with the width value. The corresponding results for both dispersions with three different thicknesses each (see number of coated layers in legend) are displayed in **Figure S9**.

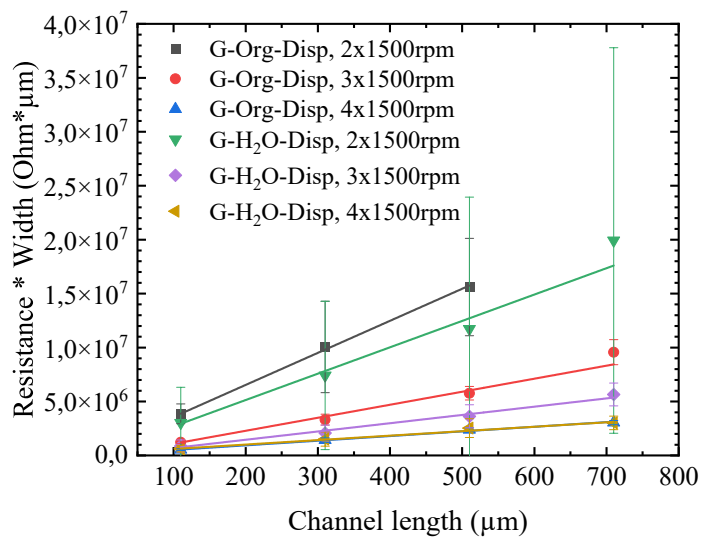

**Figure S9**: Resistance multiplied by width (due to different W considered) as a function of channel length from long-range transmission line measurements (legend: number of coated layers). The linearity of the measured resistances with respect to channel length holds in spite of channel lengths far beyond flake nominal size. Also, the fit quality increases (narrower error bars) with thicker layers, thus showing a more uniform film as discussed in the *Results and Discussion* section. The reproducibility of sheet resistance regardless of channel width

constitutes a further evidence of film uniformity and absence of localized coffee-ring-like[1] conductive paths.

**Comparison between short-range TLM and van der Pauw structures with G-H$_2$O-Disp**

The **short-range TLM** structures fabricated comprise five different channel lengths in ranges between 7μm and 210μm. A total of 15 TLM structures per number of spin coat (SC) runs were fabricated with the selected dispersion G-H$_2$O-Disp. The five different channel lengths were combined in 3 different configurations in order to increase the number of data points. Furthermore, **van der Pauw (vdP)** crosses as displayed in the inset of Fig.6a were fabricated. We compared the sheet resistances (R$_{sheet}$) acquired by both methods. The TLM measurements were performed as explained above. One channel width W=300μm (Fig. S2a) was evaluated. Fig. S2b displays the resistance values as a function of channel length for all the analyzed samples. Furthermore, we extracted the sheet resistance from the vdP measurements (6 structures per number of SC runs) according to eq. 2 (*Sheet resistance and humidity measurements* subsection of the main paper). Fig. S3 provides a comparative box plot of measured R$_{sheet}$ by means of both methods.

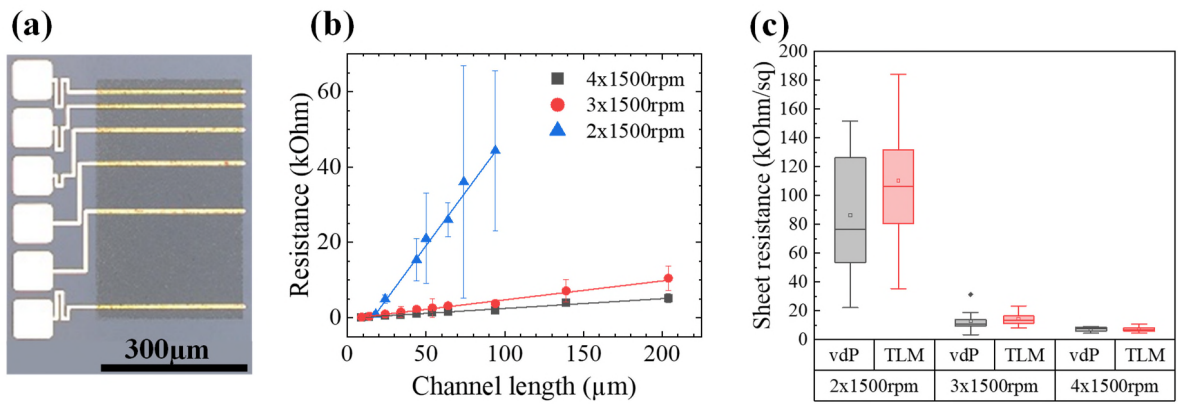

**Figure S10**: TLM measurements with G-H$_2$O-Disp and comparison with van der Pauw measurements. a) Micrograph of lift-off structured TLM structure (4x1500rpm) with five different channel lengths and W = 300μm. b) Resistance as a function of channel length for different thicknesses. Missing points on 2x1500rpm correspond to non-conductive channels. The linearity of resistances decisively improves with increased thickness, as already observed in Figure S9 and discussed in the *Results and Discussion* section. c) Comparison of measured sheet resistance via TLM and vdP measurement techniques. TLM and vdP values are in good agreement for all three thicknesses. This highlights the possibility to achieve a reproducible lift-off structured film regardless of device geometry.

### Humidity sensing with both dispersions using TLM-devices

We performed a humidity measurement similarly to Fig. 6c (albeit using different kind of devices) with both dispersions to confirm the better suitability of G-$H_2O$-Disp than G-Org-Disp as active material for humidity sensing. For this, resistance measurements on three out of the five pairs of contacts of a selected short-range TLM were simultaneously taken. We select two structures (1 per dispersion) with sheet resistances of similar value and, furthermore, with a comparable range to the ones presented with vdP structures (4x1500rpm) in Fig. 6. The measurement implementation with regard to the heating step, setting of relative humidity (RH) level and reference measurement corresponds to the description in the *Sheet resistance and humidity measurement methods* subsection. The analysis is based on two humidity pulses (set RH = 80%), the first of which is applied to the TLM device after it was exposed to a heating step to desorb water molecules from its surface[2]. The sheet resistance is calculated by extracting the slope of linear fit from the three measured resistances and multiplying it by channel width, according to eq. S2. The response is defined as $Response = (R - R_0)/R_0$, with R being the continuously measured sheet resistance $R_0$ defined as $avg(R_{sheet\_TLM})$ at RH = 17%, prior to the first humidity pulse.

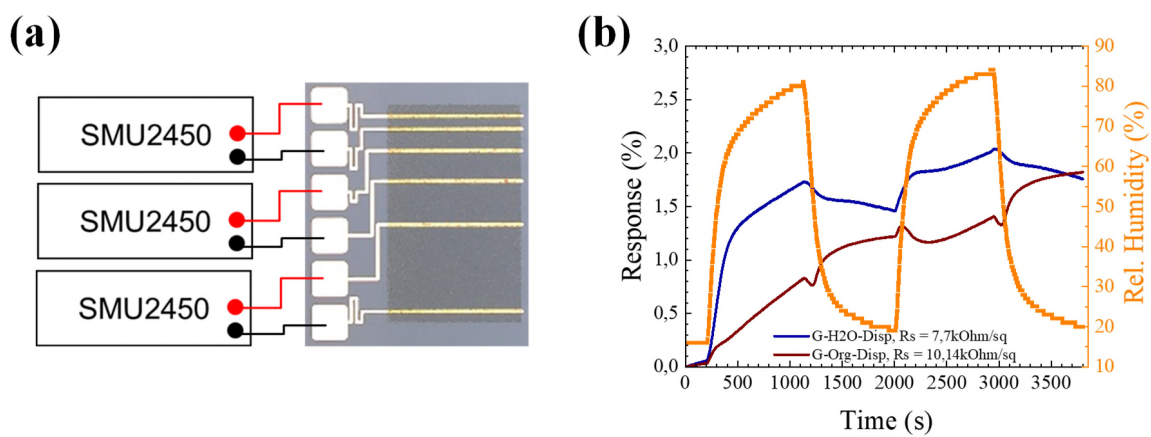

**Figure S11:** Humidity measurements on short-range-TLM structures and comparison between both dispersions. a) Illustration of resistance measurement implementation: three different pair of contacts of a same structure (1 per dispersion) were measured simultaneously and their response to two humidity pulses was extracted. b) Response of sheet resistance of one TLM structure per dispersion. The heating step enables surface cleaning by desorbing $H_2O$ molecules. This results in a higher response to a similar humidity pulse (as already reported in the *Result and Discussion* section) and can be observed with both dispersions. Moreover, G-$H_2O$-Disp shows an approximately 2 times higher response to humidity compared to G-Org-Disp when the measurement occurs upon the heating step (30s at 110°C). This confirms the better suitability of G-H$_2$O-Disp for humidity sensing despite its lower surface-to-volume ratio compared to G-Org-Disp (see AFM measurements in Fig.4d and Fig.4f). We attribute it to the higher defect density and presence of O$_2$-containing functional groups observed in the Raman measurements[3–5], which is expected to enhance the adsorption of H$_2$O-molecules[2,6,7]. The response of the sensor without a prior heating step is characterized by a lower reaction in the case of G-H$_2$O-Disp, while G-Org-Disp does not follow the resistance-increasing trend seen with the first pulse and exhibits rather an ambiguous response.

**Humidity sensing with G-H$_2$O-Disp using vdP-devices**

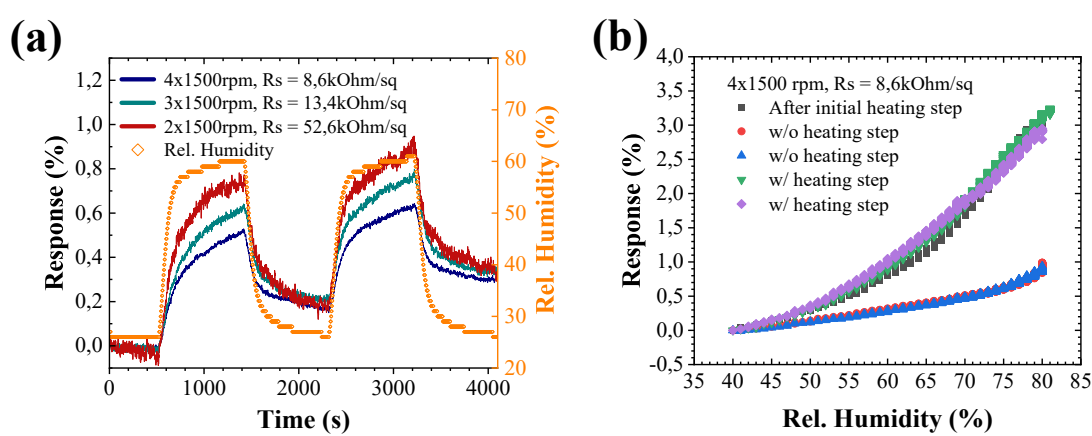

**Figure S12**: Humidity measurements of van der Pauw devices with different thicknesses and effect of heating effect on humidity sensing. a) Detail of Fig.6a: sheet resistance of all devices increases proportionally to relative humidity. The film thicknesses and sensor response are inversely correlated. Furthermore, the raw signals make clear how thinner devices lead to noisier signals. The reasons for both observed effects are discussed in the Results and Discussion section. b) Assessment of the effect of H$_2$O-desorption from the surface on humidity sensing and reproducibility of sensitivity. The inclusion of a heating step renders the sensor around 5 times more sensitive compared to the case without a heating step. The response of the sensor is reproducible throughout the experiment, which consists of three successive humidity pulses with a 5-minute heating step at 110°C in between. The reproducibility is visible through the overlapping response curves w/heating step versus relative humidity. A similarly reproducible response is observed in the curves from our experiment without the heating step. Here, we exposed the sensor to two humidity pulses without a previous heating step.